\documentclass[twocolumn,showpacs,preprintnumbers,amsmath,amssymb]{revtex4}
\usepackage{graphicx}% Include figure files
\usepackage{dcolumn}% Align table columns on decimal point
\usepackage{bm}% bold math

\begin{document}

\title{Microwave driven arbitrary coupling between trapped charge resonances in a silicon single electron transistor}% Force line breaks with \\

\author{Morteza Erfani}
 \altaffiliation{Email address: me287@cam.ac.uk}

\author{David G. Hasko}

\author{Alessandro Rossi}

\affiliation{Microelectronics Research Centre, Cavendish Laboratory, University of Cambridge, J. J. Thomson Avenue, CB3 0HE, Cambridge, United Kingdom}

\author{Wan Sik Cho}

\author{Jung-Bum Choi}

\affiliation{Department of Physics and Institute for NanoScience and Technology, Chungbuk National University, Cheongju 361-763, South Korea}

\begin{abstract}

In quantum computation, information is processed by gates that must coherently couple separate qubits. In many systems the qubits are naturally coupled, but such an always-on interaction limits the algorithms that may be implemented. Coupling interactions may also be directed in devices and circuits that are provided with additional control wiring.  This can be achieved by adjusting the gate voltage in a semiconductor device or an additional flux in a superconducting device. Such control signals must be applied adiabatically (limiting the speed) and the additional wiring provides pathways for noise, which leads to decoherence. Here we demonstrate an alternative approach to coupling by exploiting the nonlinear behaviour of a degenerately doped silicon transistor. A single transistor can exhibit a large number of individual resonances, which are seen as changes in the source-drain current of a dc-biased device. These resonances may be addressed in frequency space due to their high quality factors. Two widely separated resonances are addressed and coupled in three-frequency spectroscopy by ensuring that the third frequency corresponds to the difference between the two individual resonances. The nonlinearity causes the generation of additional driving signals with appropriate frequency and phase relationships to ensure coupling, resulting in additional spectroscopic features that could be exploited for rapid state manipulation and gate operations.

\end{abstract}

\pacs{03.67.Lx, 71.30.+h, 71.55.Gs, 72.10.Fk, 72.15.Rn, 72.20.Ee, 73.20.Fz, 73.23.Hk}
                          
\keywords{single electron transistor, silicon, microwaves, quantum information processing}

\maketitle

Qubit-qubit coupling underlies gate operation in quantum information processing (QIP) applications. Nuclear magnetic resonance (NMR) based QIP schemes rely on an always-on interaction (permanent coupling), which requires additional pulses to remove the effects of coupling, when the qubits are desired to operate independently. Superconducting quantum interference device (SQUID) based QIP circuits have also exploited permanent capacitive or inductive coupling strategies. Superconducting devices allow external control of the quantum inductance, so enabling the coupling between superconducting qubits to be controlled, as in the experimental demonstrations by Hime \textsl{et al.} \cite{Hime} and van der Ploeg \textsl{et al.} \cite{van der Ploeg}. However, this additional flexibility is limited to 1-D or 2-D circuit architectures, so that coupling between arbitrary qubit pairs can require many additional operations before the interaction can take place. Furthermore, quantum computers must enable qubit-qubit interactions to occur in the strong coupling regime in order to preserve the quantum information, i.e. the coupling time must be short compared to the phase coherence time. The phase coherence time is reduced by the addition of the extra wires necessary to control the qubit coupling, especially if these wires carry high speed signals in order to switch the coupling at high speed. By contrast, in NMR based QIP schemes, gate operations use indirectly coupled microwave or RF signals, which are globally broadcast to all qubits, but affect only those qubits that are resonant with the frequencies employed. The global broadcast of the signals that manipulate the qubit-qubit coupling avoids the need for the complex additional wiring that would lead to significant decoherence.

\begin{figure*}
\includegraphics{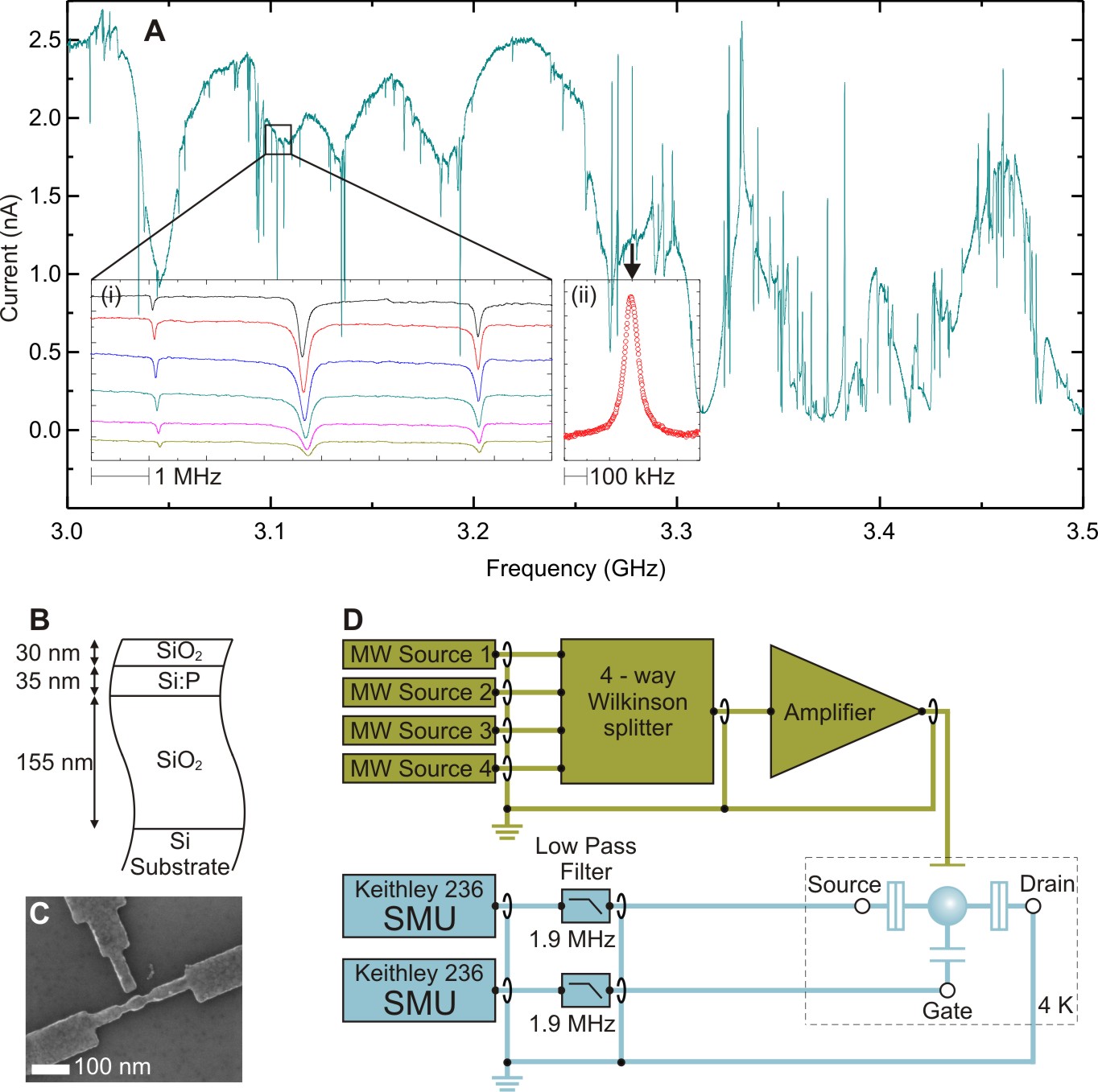}
\caption{\label{fig:fig1} \textbf{A} Single frequency CW spectroscopy on a fixed DC biased SET showing a large number of high quality factor resonances. Inset (i) shows effect of driving power alternation on three resonances where a weak dependence for the centre frequency and a strong dependence for the amplitude observed. Inset (ii) shows a typical resonance (indicated in the main diagram by the arrow) with quality factor $\sim10^5$ in a high resolution scan. \textbf{B} Schematic cross-section of the SOI substrate. \textbf{C} SEM image of the SET device. \textbf{D} Measurement setup: up to four microwave signals can be summed by a Wilkinson combiner and amplified before propagation to the SET device. DC voltage sources and current measurements are made using Keithley 236 source-measure units, with in-line low pass filters.}
\end{figure*}

Bertet \textsl{et al.} \cite{Bertet} have proposed a scheme to couple two SQUIDs with unequal energy splittings by using a microwave signal corresponding to the sum or difference between their resonant frequencies. Such a driving signal can be switched at high speed and so this approach offers a very effective method for controllable coupling. This analysis was extended by Paraoanu \cite{Paraoanu} to off-resonance microwave signals and the influence of the device non-linearity on the nature of microwave driven coupling was discussed by Liu \textsl{et al.} \cite{Liu}. In this paper we discuss an alternative approach to the coupling of coherent excitations within a degenerately doped silicon single electron transistor (SET). Previous continuous wave (CW) and pulsed microwave spectroscopy investigations have shown that even a single island SET has a rich behaviour, with many high quality factor resonances in the source$-$drain current (Fig.\ref{fig:fig1}A). The existence of both positive and negative current resonances eliminates heating effects as the origin for these resonances. Furthermore, the very large number of the resonances ( $\gg 10^4$  up to 10 GHz), and their high values for the quality factors ($\sim 10^5$), suggests that their origin is due to the oscillation of trapped electrons in the SET. Such trapping might occur if the transport is strongly localised and if coupling to the environment is weak; these conditions can be met in degenerately doped silicon if strongly depleted.  Pulsed microwave spectroscopy has previously shown Rabi-like oscillations indicating the quantum mechanical nature of these excitations \cite{Creswell1}.

Coupling between separate systems undergoing simple harmonic motion is normally encountered only when their natural frequencies are similar.  This ensures that the necessary synchronization between the systems can arise naturally.  Under special circumstances, this effect can also occur when the two frequencies are different, but take on a simple ratio.  However, if synchronization occurred as a consequence of the method of driving the system, then coupling between irrationally related frequencies can also occur.  This effect has been observed in a driven coupled nanomechanical system, consisting of two parallel conducting doubly clamped beams (mechanically joined at their centres, but otherwise electrically isolated) \cite{Shim}. One beam was driven by an ac current in a large magnetic field to excite mechanical oscillations and the other beam indicated the coupled response through induced voltage differences. These coupled responses were measured at frequencies well separated from the driving frequency. For example, Fig.S4 in their supporting online material shows coupled responses at $\sim$15.23 MHz for a driving frequency of $\sim$8.1 MHz. The frequency of the response is not a simple function of the driving frequency and also depends on the driving power (Arnold$'$s tongues), with a wide range of dependencies. The authors draw a parallel between this experimental behaviour and that observed in the theoretical predictions for an Integrate-and-fire model with coupled limit-cycle oscillators. Despite the differences between the physical mechanisms for the Integrate-and-fire model and the nanomechanical investigated experimentally, the Arnold$'$s tongue features appear characteristically similar in the two cases, suggesting a common underlying origin for the synchronization of coupled modes at different frequencies.  Such synchronization is known to affect the behaviour of arrays of quantum mechanical systems, for example Wiesenfeld \textsl{et al.} have discussed the frequency locking behaviour dynamics of a disordered array of Josephson junctions in terms of the Kuramoto model \cite{Wiesenfeld}.

In previous single frequency CW microwave spectroscopic measurements on degenerately doped SETs, the resonant frequencies are also irrationally related and must result either from single oscillating electrons behaving independently or from multiple electrons that meet the stringent set of conditions necessary for inherent coupling.  The conditions necessary for multiple electrons to be inherently coupled include; (i) their independent resonant frequencies must be very similar, (ii) the electric dipole moments due to the transferred charge need to be non-orthogonal. Clearly this first requirement greatly limits the number of possible coupled modes since the resonances are spread over a very wide frequency range.  In this work, we use multi-frequency microwave spectroscopy to explore the possibility of driven coupling following the routes suggested for superconducting devices.  Using driving frequencies that only correspond to resonances result in independent excitations; this is equivalent to driving a classical system at a mode frequency.  But driving at a frequency difference between a pair of resonances can be used to cause synchronization between otherwise independent excitations.  The coupling that results from this synchronization is seen in the appearance of new spectroscopic features.  The amplitude and location in frequency of these new features is used to characterise the coupling strength.  The relevance for this directed coupling technique to quantum information processing is discussed.

SETs were fabricated on degenerately doped silicon-on-insulator wafers (see schematic cross-section of the wafer in Fig.\ref{fig:fig1}B) using high-resolution electron beam lithography and reactive ion etching \cite{Ali}, see SEM image in Fig.\ref{fig:fig1}C.  All experiments were performed with fixed source$-$drain and gate bias voltages with the device maintained at 4K by direct immersion in liquid helium.  DC signals were applied directly from Keithley 236 source-measure units, low pass filtered at room temperature in order to avoid sample heating by electrical noise, see Fig.\ref{fig:fig1}D.  Microwave signals were applied indirectly using a low-Q cavity formed by the space ($\sim$0.5 mm) between an open-ended semi-rigid coaxial cable and the header package supporting the device-under-test.

The microwave signals were provided by multiple AtlanTecRF ANS3 programmable sources; the individual signals were summed using a Wilkinson combiner, then amplified before delivery to the low-Q sample cavity.  This microwave delivery system does not modify the spectral content provided that the combiner and power amplifier have linear behaviour and that a large separation between the individual frequencies is maintained. The first requirement is met by limiting the input to the power amplifier to less than +7 dBm. This is somewhat below the manufacturers recommended maximum input, but measurements indicate the onset of saturation at this value (albeit at a very low level of distortion). The second requirement results from the need to avoid phase locking between the microwave signal sources; which causes the actual output frequency to deviate from the desired frequency if the two desired frequencies approach too closely. It has been observed when the two frequencies are closer than $\sim$40 kHz, frequency locking between the two sources take place. In this work, a wide separation in frequencies is maintained and simple attenuators can be used to isolate sources; if closer separation in frequencies is required, sources could be isolated by circulators to avoid this problem.

\begin{figure}
\includegraphics{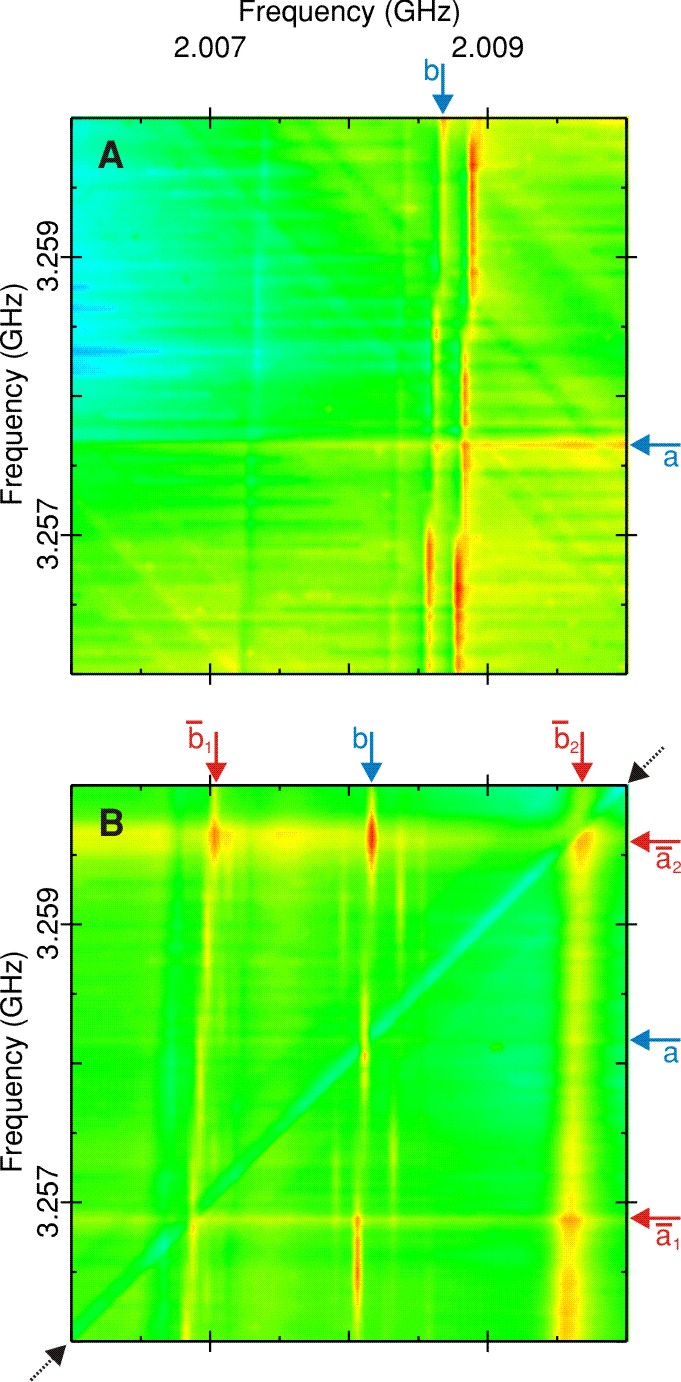}
\caption{\label{fig:fig2} \textbf{A} 2-D microwave spectroscopy showing two prominent resonances indicated by the $a$ and $b$ arrows. \textbf{B} 3-D microwave spectroscopy with $f_z =$1.25002 GHz showing additional features indicated by the $\overline{a}$ and $\overline{b}$ arrows. The black dotted arrows indicate the trace corresponding to the difference of $f_x$ and $f_y$ being equal to $f_z$.}
\end{figure}

\begin{figure*}
\includegraphics{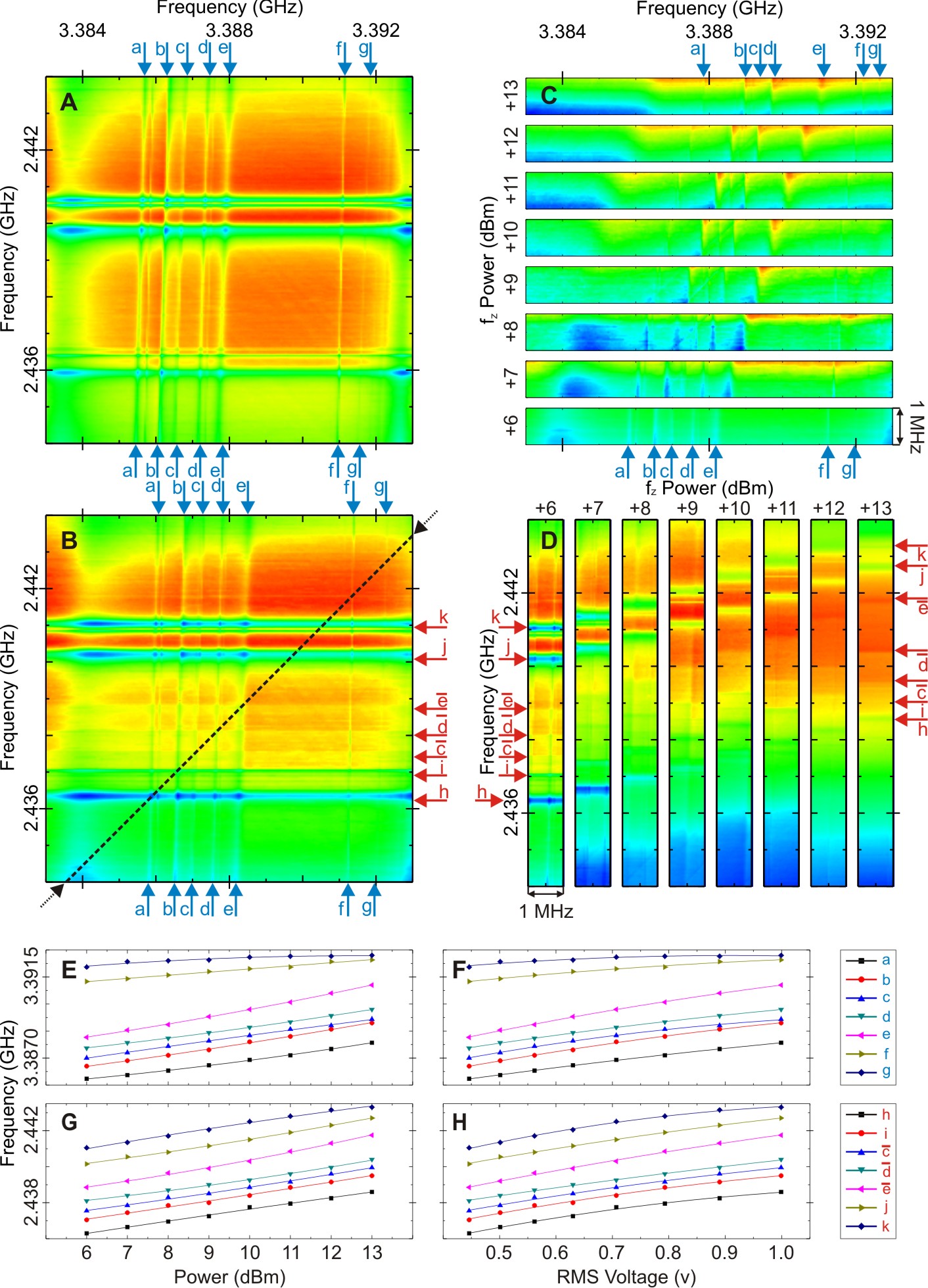}
\caption{\label{fig:fig3} \textbf{A} 2-D microwave spectroscopy and \textbf{B} 3-D microwave spectroscopy with $f_z =$0.94955 GHz. \textbf{C} and \textbf{D} limited range 3-D microwave spectroscopy with the $f_z$ power stepped at +1 dBm intervals showing a systematic change in the centre frequency of resonances in the $f_x$ and $f_y$ ranges respectively. \textbf{E} and \textbf{G} show the positions of selected resonances in $f_x$ and $f_y$ with respect to the $f_z$ power.  \textbf{F} and \textbf{H} show the positions of selected resonances in $f_x$ and $f_y$ with respect to the $f_z$ RMS voltage amplitude.}
\end{figure*}

Single frequency spectroscopy (Fig.\ref{fig:fig1}A) shows a large number of high quality factor resonances, so that to avoid confusion in multi-frequency spectroscopy we must greatly restrict the measured frequency ranges as shown in inset (i) of Fig.\ref{fig:fig1}A. Fig.\ref{fig:fig2}A shows the change in SET current in a colour plot map in response to the simultaneous application of two microwave signals;  $f_x$ was swept in a narrow range close to 2.00 GHz, while $f_y$  was stepped in a similar range close to 3.25 GHz.  Two prominent resonances, labelled $a$ and $b$, are highlighted by arrows on this figure; the very nearly horizontal and vertical character of these features indicates their independence, even where the resonances intersect.  The much fainter background features, with -45 degree slope, are due to resonances in the range of 5.262 to 5.272 GHz, which are visible due to mixing of the driving signals $f_x$ and $f_y$ by the non-linearity of the SET.  The very small $RC$ time constant, due to the island capacitance and tunnel barrier resistance, allows this process to occur up to very high frequencies, but at very low efficiency, so that the amplitude of the resonance is small.   Both sum and difference frequencies are generated, but in this case the background features correspond to the former.

Applying a third microwave signal, with $f_z$ fixed at 1.25002 GHz, significantly alters the SET response, see Fig.\ref{fig:fig2}B.  Two new resonances appear on each axis, as shown by the $\overline{a}$ and $\overline{b}$ arrows, along with a small shift in the frequencies of the original resonances $a$ and $b$, as well as significant changes in amplitude.  The spacing between the $\overline{a}$ resonances is the same as the spacing between the $\overline{b}$ resonances, at about 3 MHz. In addition, these new resonances cross at intersections with a line at +45 degrees slope exhibiting a fainter background feature that corresponds to the difference between $f_x$ and $f_y$  being equal to $f_z$ , shown by dotted arrows. This suggests that the new resonances have a common origin associated with the third frequency and the mixing process.  If the two independent resonances seen in Fig.\ref{fig:fig2}A were coupled, then we expect to see two new resonances corresponding to the symmetric and anti-symmetric modes. Such a coupling could only be realized at very different frequencies if there were some mechanism to synchronize their motion.  The mixing process creates a signal in the $f_x$ frequency range using one in the $f_y$ frequency range combined with the $f_z$ signal, and similarly with $f_x$ and $f_y$ reversed.  Clearly, these two mixed products are related by the common $f_z$ signal and it is this relationship that allows the oscillations in the two frequency ranges, $f_x$ and $f_y$, to become synchronized.  However, since mixing by the device non-linearity is very inefficient these new coupled mode resonances would be expected to be rather faint. If the mixing were to take place within the oscillating electron systems giving rise to the resonances, then this effect would be significantly enhanced.  Assuming the higher frequency resonance to correspond to the anti-symmetric mode in both frequency ranges and the lower frequency resonance to correspond to the symmetric mode, then their superposition behaviour can be understood.  Superposing both the anti-symmetric and symmetric modes results in resonance addition (the top left and bottom right crossings); but superposing the same mode, either the anti-symettric or the symmetric mode (the bottom left and top right crossings) results in resonance subtraction, in Fig.\ref{fig:fig2}B. The mode splitting of 3 MHz implies an energy exchange time of $\sim3$  $\mu s$, corresponding to a 2$\pi$ pulse; this is sufficiently small compared to the resonance lifetime that a large number of operations could be completed within the resonance lifetime.

The magnitude of the mode splitting is a measure of the coupling strength and so should depend on the strength of the mixed products.  Fig.\ref{fig:fig3} shows a systematic investigation of the dependence of the frequency of some coupled and un-coupled resonances on the power applied at $f_z$.  Fig.\ref{fig:fig3}A and \ref{fig:fig3}B show colour plot maps of the behaviour without and with the application of $f_z$; as before we see new peaks and shifts in the positions of the original resonances.  The effect of changing the power of $f_z$ is shown in Fig.\ref{fig:fig3}C and \ref{fig:fig3}D, where the position of resonances is indicated in the lowest 1 MHz frequency range in $f_x$ and $f_y$ respectively.  Systematic changes in peak position are seen as the power of $f_z$ is increased by +1 dBm steps. The behaviour of selected resonances in $x$-axis is shown in Fig.\ref{fig:fig3}E and \ref{fig:fig3}F as a function of driving power and driving root mean square (RMS) voltage. Similarly, the behaviour of selected resonances in $y$-axis is shown in Fig.\ref{fig:fig3}G and \ref{fig:fig3}H. Two types of behaviour can be identified; the upper two traces in Fig.\ref{fig:fig3}E show little variation with the power of the $f_z$ driving signal and may be regarded as uncoupled, whereas the remaining traces show an approximately linear increase in frequency with $f_z$ power.  Previous work has shown that single frequency excitation does not significantly change the resonant frequency with changing power \cite{Creswell2}, see inset (i) of Fig.\ref{fig:fig1}A. 

The coupled modes in multi-coordinate systems may be found by analysis of the differential equations \cite{Main}, and shows that the mode frequencies of asymmetric systems increase as the coupling strength increases.  The SET island is expected to contain a significant number of localized electrons; any particular mode may involve one or more of these electrons resulting in a very much larger number of resonances.  Any such oscillation results in a time dependent dipole (or more complex) electric field that can in principle affect any other system.  However, any localized electron that is not oscillating will show no dipole field (probably only a very short range radial field centred on the trap positive charge) and so cannot easily interact with other systems.  Moreover, for synchronization to occur in undriven systems, it is necessary for the resonant frequencies to be a simple multiple of each other (usually one to one).  These limitations are overcome by the strategy of driving the system at the frequency difference between resonances.  In this case, resonances can be excited at widely separated frequencies, but still be effectively synchronized by the common driving signal.  The strength of this common signal plays the same role as the stiffness of the coupling spring in a classical coupled system, but in this case the ‘stiffness’ is directly controlled by the power of the third frequency.

To exploit these effects it is necessary to initialise individual oscillators using on-resonance pulses; then interactions may be imposed between selected oscillators using pulses at their resonant frequency difference.  The nature of the interaction is determined by the strength and duration of the frequency difference pulses.  Finally, readout is achieved also with on-resonance pulses, but with the limitation that only a single parameter (bit) may be read out after a single computation.  Clearly, the number of bits is not limited by the extent of the circuit, sufficient already exist in a single transistor device.  Single bit readout is also not a significant disadvantage as long as extensive averaging is not required (all results presented here are the unaveraged data).  However, the main challenging requirement for an extensive computer is the need for large numbers of individual, and rapidly switchable, microwave sources.

%***********************************************************************

%*******************************************************************************

\end{document}